\documentclass[aip,jmp,amsmath,amssymb,reprint]{revtex4-1}
\usepackage{graphicx}
\usepackage{dcolumn}
\usepackage{bm}
\usepackage{color}
\usepackage{amsmath}
\usepackage{wasysym}
\usepackage{amssymb}

\begin{document}
\preprint{AIP/123-QED}

\title{Spatial anisotropy and heterogeneity in contractility and adhesion
  distribution may contribute to cell steering during migration}

\author{Soumya S S}\affiliation{Department of Civil Engineering,
 Indian Institute of Technology Bombay, Mumbai 400076, India}
\author{Subodh Kolwankar}\affiliation{Department of Civil Engineering,
Indian Institute of Technology Bombay, Mumbai 400076, India}
\author{Edna George}\affiliation{WRCBB,
 Department of Biosciences and Bioengineering,
 Indian Institute of Technology Bombay, Mumbai 400076 India}
\author{Santanu K. Basu}\affiliation{Department of Chemical Engineering,
 Indian Institute of Technology Bombay, Mumbai 400076 India}
\author{Shamik Sen}\email{shamiks@iitb.ac.in}\affiliation{WRCBB,
 Department of Biosciences and Bioengineering,
 Indian Institute of Technology Bombay, Mumbai 400076 India}
\author{Mandar M. Inamdar}\email{minamdar@civil.iitb.ac.in}
\affiliation{Department of Civil Engineering,
 Indian Institute of Technology Bombay, Mumbai 400076, India}

\date{\today}

\begin{abstract}

Transition from random to persistent cell motility requires
spatiotemporal organization of the cytoskeleton and focal adhesions.
The influence of these two structures on cell steering can also be
gleaned from trypsin de-adhesion experiments, wherein cells exposed to
trypsin round up, exhibiting a combination of rotation and
translation. Here, we present a model to evaluate the contributions of
contractility and bond distribution to experimentally observed
de-adhesion. We show that while asymmetry in bond distribution causes
only cell translation, a combination of asymmetric bond distribution
and non-uniform contractility are required for translation and
rotation, and may guide cell migration.
\end{abstract}
\maketitle

Cell motility is central to various processes including embryogenesis
and wound healing. For motility to occur, cells must possess a
polarized shape defined by leading and trailing edges, with the cell
protruding at the leading edge and retracting at the trailing
edge~\cite{Ridley2003}. While cell protrusions are stabilized by the
formation of integrin--based adhesions, cell retraction requires the
detachment of rear adhesions orchestrated by actomyosin-based
contractile forces. Efficient migration, therefore, requires close
temporal and spatial co-ordination between the cytoskeleton and the
adhesion apparatus. On two-dimensional ($2-$D) substrates, most adherent
cells exhibit random cell motility, frequently altering their leading
and trailing edges. In contrast, in the presence of
directional cues, persistent cell migration is
observed.\cite{Ryan2009} Since both the cytoskeleton and the adhesions
contribute to motility, their relative contributions to such frequent
directional changes remain unclear. Moreover, because most mesenchymal
cells possess speeds in the order of few microns/hour, for motility to
occur, cell shape stability must be maintained at shorter timescales
(few minutes) via the force balance between contractile actomyosin
forces and the focal adhesions that withstand these forces. This tensional homeostasis can be gauged using the trypsin de-adhesion assay wherein, cell retraction is tracked upon
rapidly severing cell-matrix contacts using the enzyme trypsin
(Fig.~1 and corresponding Supp. Movie)~\cite{sen2009}. Further, sigmoidal retraction kinetics
with two time constants observed for a wide variety of cells, including
fibroblasts, epithelial cells and cancer cells, is indicative of the
generality of the process (Fig.~1b).

\begin{figure}
\includegraphics[width=8.5cm]{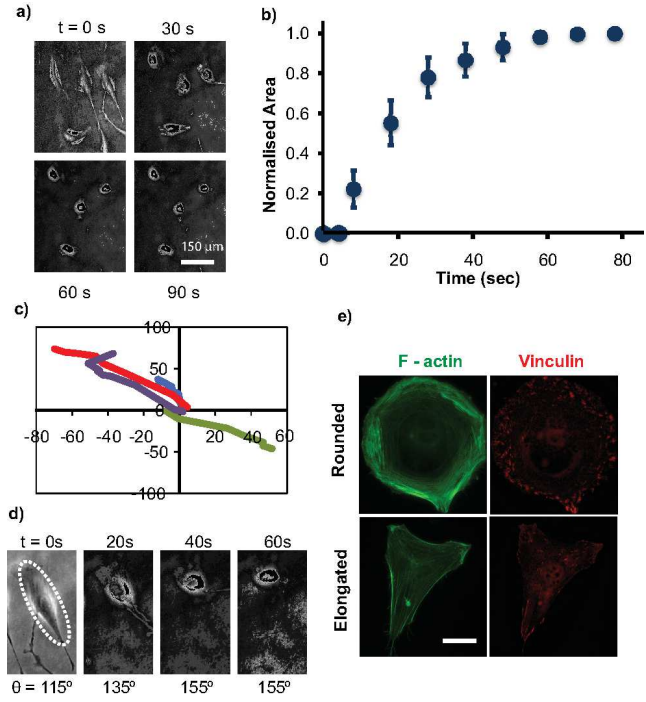}
\caption{\label{fig:1} (Color online) Trypsin induced de-adhesion of
  NIH 3T3 fibroblasts. Fibroblasts were cultured on collagen-coated
  substrates for 24 hours. Before experiment, cells were washed with
  PBS, warm trypsin was added and images were acquired in time-lapse
  till cells rounded up but remained attached to the substrate. Also see the corresponding supplementary movie. (a) Sequence of time-lapse images of fibroblasts rounding up upon
  addition of trypsin. (b) Plot of normalized area (defined as change
  in area at any time divided by net change in area during the entire
  de-adhesion process) as a function of time. (c) During de-adhesion,
  cells underwent both translation and rotation. Quantification of
  translation obtained by tracking the centroid of individual cells
  during the time course of de-adhesion. (d) Sequence of images of a
  cell undergoing rotation during de-adhesion. At each time point, the
  cell is approximated by an ellipse, with θ indicating the
  orientation of the major axis of the cell with the x-axis. (e)
  Cytoskeletal and focal adhesion organization in cells with round and
  elongated morphology. Cells were fixed and stained for F-actin
  (green) for visualizing the actin cytoskeleton and with vinculin
  (red) for observing the distribution of focal adhesions (enhanced online).}
\end{figure}

When adherent 3T3 fibroblasts were incubated with the enzyme trypsin,
cells exhibited a combination of translation and rotation during the
de-adhesion (or rounding up) process (Figs.~\ref{fig:1}c~and~\ref{fig:1}d ). Such motions
are probably indicative of a combination of anisotropy in cytoskeletal
contractile forces (also referred to as cell prestress) and in cell-matrix
adhesions (Fig.~\ref{fig:1}e). If such anisotropy exist, this may help us to understand
the frequent directional changes associated with random cell motility.
In this letter, we specifically address this question by studying the
crosstalk between distributions of cell prestress and adhesion on the
pattern of de-adhesion exhibited by 3T3 fibroblasts. Here, we propose a
simple theoretical model which can replicate the above mentioned cell
dynamics during the processes of  de-adhesion.

  To understand the role of various factors dictating de-adhesion
  patterns (Fig.~\ref{fig:1} and supplementary information (SI) Fig.~S1), we began with a $2-$D
  continuum description of a prestressed cell in mechanical
  equilibrium with its substrate.  The cell was modeled as a circle of
  radius $R_{\rm cell}$ connected to the substrate by continuous bonds
  -- both cell and the substrate were represented as continuous,
  homogeneous, isotropic visco-elastic Kelvin-Voigt
  materials\cite{Mesquita2003,Karcher2003}.  The cell-substrate bonds
  were modeled using linear springs~\cite{schwarz2006} whose number
  and distribution depended on the type of distribution of focal
  adhesion between cell and substrate. Since the thickness of the cell
  is typically small as compared to its other dimensions, it was
  reasonable to assume condition of plane stress for the present
  problem\cite{edwards2011,banerjee2012,Banerjee2013}.
  \begin{figure}
\includegraphics[width=8.5cm]{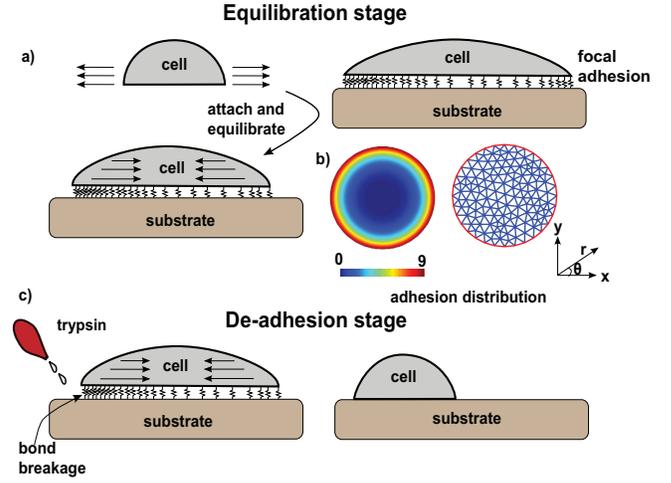}
\caption{\label{fig:2}(Color online) A schematic representation of our
  modeling of cell adhesion and de-adhesion dynamics. (a) Circular
  cell was subjected to stretching to mimic contractility. Next, the
  stretched cell adhered to substrate to represent formation of focal
  adhesions. Cell then reached a mechanical equilibrium state or
  tensional homeostasis with substrate by balancing the contractile
  forces generated by actin-myosin cytoskeleton (inward arrows) and
  cell-substrate adhesions (springs). (b) To simulate the de-adhesion
  dynamics of the cell, a finite element analysis was performed. For
  that, entire domain was discretized into triangular elements. The
  non uniform distribution of focal adhesion was modeled by varying
  the number of connected springs over the cell area. (c) De-adhesion
  process was initiated by adding trypsin to the cell in equilibrium.
  Upon treatment with trypsin, the focal adhesions were broken in a
  time dependent manner and cell started to relax. When de-adhesion
  process was completed, all the focal adhesions were broken and cell
  attained its original circular shape from the stretched
  configuration.}
\end{figure}

Figures \ref{fig:2}a and \ref{fig:2}c schematically describe the
adhesion and de-adhesion dynamics of the cell. For the purpose of a
systematic study, the entire sequence of cell activity was divided
into two stages; the initial equilibration stage and the subsequent
de-adhesion stage. During the first stage or equilibration stage, the
cell which had been initially in a relaxed round shape was brought to
a stretched state by the application of specified displacement fields
$u_0(x,y)$ and $v_0(x,y)$ (equivalent to ``thermal''
prestress~\cite{edwards2011}) and then attached to the substrate with
the help of cell-substrate bonds. Once the cell was connected to the
substrate, it started applying tractions on the substrate and
ultimately reached a mechanical equilibrium state with the substrate
as shown in Fig.~\ref{fig:2}a. At this stage, the equation of
equilibrium of the system, in terms of displacement components of the
cell, $u_{c}$ and $v_{c}$, was
\begin{eqnarray}
\frac{1}{2\left(1+\nu\right)}\nabla^{2}u_{c}
+\frac{1}{2\left(1-\nu\right)}\frac{\partial}{\partial x}\left(\frac{\partial u_{c}}{\partial x}
+\frac{\partial v_{c}}{\partial y}\right)\nonumber\\
+r_{1}\rho_{0}\left(x,y\right)\left(u_{0}-u_{c}\right)=0,
\label{eq:equilib1}
\end{eqnarray}
\begin{eqnarray}
\frac{1}{2\left(1+\nu\right)}\nabla^{2}v_{c}
+\frac{1}{2\left(1-\nu\right)}\frac{\partial}{\partial y}\left(\frac{\partial u_{c}}{\partial x}
+\frac{\partial v_{c}}{\partial y}\right)\nonumber\\
+r_{1}\rho_{0}\left(x,y\right)\left(v_{0}-v_{c}\right)=0,
\label{eq:equilib2}
\end{eqnarray}
where $r_1$ represents ratio of {\it effective} substrate stiffness
with respect to {\it effective} cell stiffness (see SI Sec.~S2 for
derivation), and $\nu$ ($= 0.45$ for near incompressibility) is the
Poisson's ratio of the cell. The density of adhesion over cell area is
denoted by $\rho_{0}\left(x,y\right)$. Figure~\ref{fig:2}b shows a
qualitative axially symmetric distribution of adhesion bonds over cell
area. The value of cell displacements, $u_{c}$ and $v_{c}$ at
equilibrium state (time $t=0$) were obtained by solving
Eqs.~\ref{eq:equilib1} and~\ref{eq:equilib2} numerically by finite
element analysis (FEA) using PDE toolbox of \emph{Matlab} with stress
free boundary conditions at the edge. Note that we have
non-dimensionalized all length-units with respect to $R_{\rm cell}$
for compactness of equations (see SI for additional details).

The second stage or the de-adhesion stage was modeled by assuming a
force independent first order kinetics with rate $r$ for bond
breakage, beginning from the equilibrium configuration generated above,
such that
$\rho_{0}\left(x,y,t\right)=\rho_{0}\left(x,y\right)\exp\left(-rt\right)$
was the bond density function at any time $t$ (Fig.~\ref{fig:2}c). The resulting
non-dimensionalized equation of motion was
\begin{widetext}
\begin{eqnarray}
\frac{1}{2\left(1+\nu\right)}\left(\nabla^{2}u+\frac{\partial}{\partial{t}}\nabla^{2}u\right)
+\frac{1}{2\left(1-\nu\right)}\frac{\partial}{\partial x}\left(\frac{\partial u}{\partial x}
+\frac{\partial v}{\partial y}\right)
+\frac{1}{2\left(1-\nu\right)}\frac{\partial}{\partial{t}}\frac{\partial}{\partial x}\left(\frac{\partial u}{\partial x}
+\frac{\partial v}{\partial y}\right)\nonumber\\
+\rho_{0}\left(x,y\right)\exp\left(-r_{2}{t}\right)\left(r_{1}\left(u_{0}-u\right)-r_{3}\frac{\partial u}{\partial{t}}\right)=0,
\label{eq:time1}
\end{eqnarray}
\begin{eqnarray}
\frac{1}{2\left(1+\nu\right)}\left(\nabla^{2}v+\frac{\partial}{\partial{t}}\nabla^{2}v\right)+
\frac{1}{2\left(1-\nu\right)}\frac{\partial}{\partial y}\left(\frac{\partial u}{\partial x}
+\frac{\partial v}{\partial y}\right)+\frac{1}{2\left(1-\nu\right)}\frac{\partial}{\partial{t}}\frac{\partial}{\partial y}\left(\frac{\partial u}{\partial x}+\frac{\partial v}{\partial y}\right)\nonumber\\
+\rho_{0}\left(x,y\right)\exp\left(-r_{2}{t}\right)\left(r_{1}\left(v_{0}-v\right)-r_{3}\frac{\partial v}{\partial{t}}\right)=0.
\label{eq:time2}
\end{eqnarray}
\end{widetext}
The non-dimensionalization of all time-units was done with respect to
$\tau_0$, which is the ratio of cell viscosity with cell stiffness. We
also introduced two additional parameters $r_{2}$ --
non-dimensionalized bond cutting rate -- and $r_{3}$ -- the ratio of
effective substrate viscosity to cell viscosity in these equations
(see SI Sec.~S2). The length scale for non-dimensionalization was
$R_{\rm cell}$ as before.

Taking the values of cell displacement at equilibrium state ($u_{c}$
and $v_{c}$) from the solution of
Eqs.~\ref{eq:equilib1}~and~\ref{eq:equilib2} as the initial
conditions, time dependent PDEs Eqs.~\ref{eq:time1}~and~\ref{eq:time2}
were solved to get the cell displacement values, $u(x,y,t)$ and $v(x,y,t)$, at any time $t$. To
solve Eqs.~\ref{eq:time1} and \ref{eq:time2} numerically, we again
used \emph{Matlab} PDE toolbox as earlier with \texttt{ode15s} solver for
time integration. A sample discretized mesh used for the FEA is shown
in Fig.~\ref{fig:2}b.

Our model could recover the sigmoidal behaviour for the dynamics of
projected cell area as shown in Fig.~1b (see SI Fig.~S2). We then
focused on the theoretical understanding of translation and rotation
motion of cell during de-adhesion process. To estimate the translation
of cell, we tracked the displacement of the cell centroid as a
function of time. The overall rotation of the cell at any time, $t$
was calculated as $\frac{\int\omega dA}{\int dA}$, where
$\omega=\frac{1}{2}\left(\frac{\partial u}{\partial y}-\frac{\partial
  v}{\partial x}\right)$ is the rotation tensor, and $u$ and $v$ the
cell displacements at the time $t$.
\begin{figure}
\includegraphics[width=8.5cm]{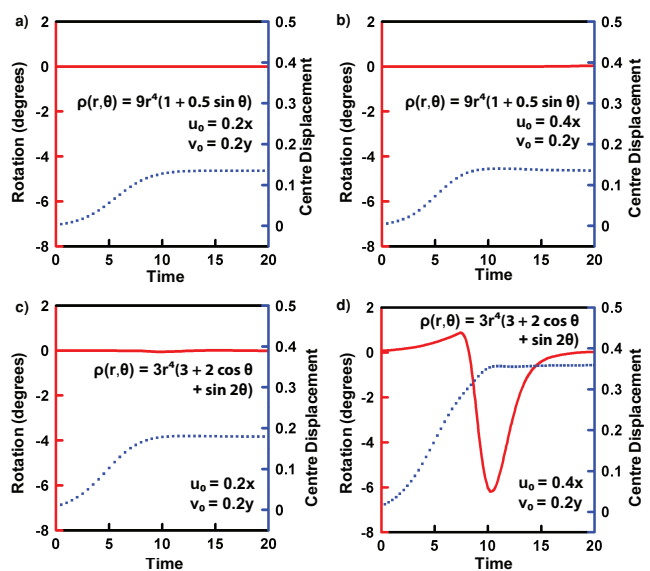}
\caption{\label{fig:3}(Color online) Effect of bond distribution and
  prestress on cell translation and rotation during de-adhesion. Plots between cell
  center displacement and rotation with time for various combination
  of bond distribution and prestress are shown. (a) Effect of axially
  asymmetric bond distribution on cell translation and rotation. (b)
  Combined effect of axially asymmetric bond distribution and
  anisotropic prestress on cell translation and rotation. (c) Effect
  of completely asymmetric bond distribution on cell translation and
  rotation. (d) Combined effect completely asymmetric bond
  distribution and anisotropic prestress on cell translation and
  rotation. For all these simulations values of non-dimensionalized
  parameters $r_{1}$, $r_{2}$ and $r_{3}$ were taken as $r_{1}=50$,
  $r_{2}=0.6$ and $r_{3}=0.1$; $\theta$ is as shown in Fig.~\ref{fig:2}b.}
\end{figure}
\begin{figure}
\includegraphics[width=8.5cm]{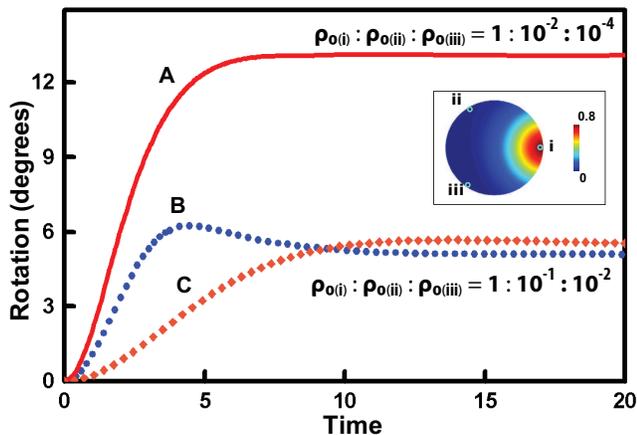}
\caption{\label{fig:4} (Color online) Role of bond distribution on
  cell rotation. (A), (B) and (C) show cell rotation with time for
  different values of bond density at three points ($i$, $ii$ and
  $iii$) on cell periphery, where the cell-substrate bonds were
  prominently concentrated. (A) Bond densities at $i$, $ii$ and $iii$
  were significantly different (of the order of 100) from each other.
  (B), (C) The bond distribution function was same for both of these
  and the bond densities at $i$, $ii$ and $iii$ were moderately
  different (of the order of 10) from each other. During the process
  of de-adhesion, bonds at all the three points were broken in a
  uniform manner for (A) and (B). For (C), only bonds concentrated at
  two points ($ii$ and $iii$) were broken and the point with highest
  bond density ($i$) remained intact (expressions for bond
  distribution function for (A), (B) and (C) are shown in
  SI Sec.~S3). Inset showing the qualitative
  representation of bond densities at points $i$, $ii$ and $iii$ for
  (A), (B) and (C). For all these simulations, values of
  non-dimensionalized parameters were taken as $r_{1}=10$, $r_{2}=0.4$
  and $r_{3}=0.1$, with $u_0=0.6x$ and $v_0=0.2y$.}
\end{figure}

We investigated the role of contractility ($u_0(x,y), v_0(x,y)$) and
bond distribution ($\rho_0(x, y)$) on cell translation and rotation
during de-adhesion. In order to understand the influence of adhesion
distributions on cell dynamics, three different types of bond
distributions were considered in the study. To begin with, a fully
symmetric adhesion distribution was applied throughout the cell
contact area. In the second case, we took a bond distribution function
with only one axis of symmetry (axially asymmetric bond distribution).
Finally, completely asymmetric bond distribution (no axis of symmetry)
was applied to the cell. In all the three cases, bond distribution was
chosen in such a manner to generate highest number of adhesions at
cell periphery and least at the centre, as is observed
experimentally\cite{Geiger2009}. Care had also been taken to keep the total
number of bonds $\int_{A} \rho_0(x,y)dA$ constant in all the cases.
Similarly, for studying the effect of the nature of prestress on cell
motions, we imposed two different types of prestress on cell;
isotropic and anisotropic. Isotropic and anisotropic prestress were
created by providing displacement fields, $u_{0}(x,y)$ and
$v_{0}(x,y)$, of equal and unequal intensities along $x$ and $y$
direction, respectively, in the equilibration stage. Various
combinations of bond distribution and prestress were applied to the
cell and its translation and rotation upon de-adhesion were examined
in each case. Throughout the simulations, the values of non
dimensional parameters, $r_{1}$, $r_{2}$ and $r_{3}$ were kept
constant.

After plotting cell centroid displacement and rotation with time, the
following observations were made. If the cell-substrate bonds were
symmetrically distributed, cell neither translated nor rotated during
the de-adhesion process, irrespective of the nature of prestress. For
an axially asymmetric distribution of bonds, translational motion
occurred for both uniform and anisotropic prestress. Here we could not
see cell rotation (Figs.~ \ref{fig:3}a and \ref{fig:3}b). Even when
the bond distribution was completely asymmetric, cell did not rotate
upon de-adhesion if the applied prestress was uniform in both
directions (Fig.~\ref{fig:3}c). Finally, when the cell-substrate bonds
were completely asymmetric and the applied prestress was non uniform,
along with pure translation, cell exhibited some steering motion also
(Fig.~\ref{fig:3}d).

Although under the influence of anisotropic prestress and complete
asymmetry in the bond distribution, the modeled cell exhibited a
combination of translation and rotation, its behaviour did not mirror
the experimental observation, where the observed rotation was largely
monotonic (Fig.~\ref{fig:1}d and Supp. Movie). Hence, as the next
case, we tried a different type of bond heterogeneity and saw its
effect on cell rotation. For that purpose, cell substrate bonds were
assumed to be predominantly concentrated at three points on the cell
periphery as opposed to the continuous of distribution of bonds in the
earlier cases (SI Sec. S3) -- prestress was also kept to be
anisotropic. Figure~\ref{fig:4} shows the plot of cell rotation with
time for three different bond densities $\rho_0$ at points $i$, $ii$
and $iii$. From the plot, it can be observed that cell with non
uniform values of bond densities at different points on its periphery
exhibited monotonic rotation during the process of de-adhesion. It
was also noted that, as the difference in bond densities between
various points increased, the amount of cell rotation also increased
(Fig.~\ref{fig:4}A versus Fig.~\ref{fig:4}B).

In conclusion, our results indicate that cellular translation arises
due to partial asymmetry about any axis in the bond distribution
and(or) bond strength, with contractility dictating the magnitude of
the final movement (Fig.~3). In contrast, cellular rotation requires a
combination of radial asymmetries in both bond distribution as well as
contractility, with sustained monotonic rotational movement requiring
a highly polarized distribution of finite adhesion spots
(Figs.~3~and~4). Since cell migration involves the co-ordinated
formation and breakage of adhesions, therefore, the spatial
heterogeneity in adhesion distribution and cytoskeletal organization
can impact both random as well as directed migration depending on the
extent of spatio-temporal coupling between these two types of
structures. Despite the complex mechanochemistry regulating cellular
movements, our simulated de-adhesion experiments and theory directly
implicate anisotropy and heterogeneity of adhesion distribution and
contractility as two of the important factors influencing the
directional changes associated with cell motility.

{\bf Acknowledgement:} MMI gratefully acknowledges financial support from Department of Science and Technology, India.
%

 \end{document}